\documentstyle[11pt,newpaspbio,twoside]{article}
\def\edcomment#1{\iffalse\marginpar{\raggedright\sl#1\/}\else\relax\fi}
\reversemarginpar
\input epsf 

\begin{document}

\title{Captain Cook, the Terrestrial Planet Finder and the Search for Extraterrestrial Intelligence}

\author{Charles Beichman}
\affil{Jet Propulsion Laboratory,
California Institute of Technology,
Pasadena, CA 91109}

\begin{abstract}

Over two hundred years ago Capt. James Cook sailed up Whitsunday Passage, just a few miles where we now sit, on a voyage of astronomical observation and discovery that remains an inspiration to us all. Since the prospects of our visiting planets beyond our solar system are slim, we will have to content ourselves with searching for life using remote sensing, not sailing ships. Fortunately, a recently completed NASA study has concluded that a Terrestrial Planet Finder could be launched within a decade to detect terrestrial planets around nearby stars. A visible light coronagraph using an 8-10 m telescope, or an infrared nulling interferometer, operated on either a $\sim40$ m structure or separated spacecraft, could survey over 150 stars, looking for habitable planets and signs of primitive life. Such a mission, complemented by projects (Kepler and Eddington) that will provide statistical information on the frequency of Earth-sized planets in the habitable zone, will determine key terms in the ``Drake equation'' that describes the number of  intelligent civilizations in the Universe.

\end{abstract}

\section{Historical Introduction}

This conference takes place in the Whitsunday group of islands discovered on Whitsunday (June 3), 1770, by the English explorer, Capt. James Cook. Surprisingly, the stated goal of Cook's voyage was astronomical in nature. As described in Richard Hough's excellent biography  (1997) of Capt. Cook, the story has echoes in today's searches for other worlds. Johannes Kepler and Edmund Halley (1716) predicted that Venus would traverse across the face of the Sun in 1761 and 1769.\footnote{Transits occur in pairs separated by roughly a century. The next pair will happen on June 8, 2004, and June 5, 2012. See {\it The Transit of Venus} (Sellers , 2001).}  These astronomers realized that comparing the timing of the transit from multiple locations would  yield a measurement of cosmological importance, the Earth-Sun distance or Astronomical Unit (AU), then known to no better than a factor of two (Sellers, 2001; and {\it www.dsellers.demon.co.uk/ venus/ven\_ch4.htm)}. Just as today's scientists argued that  the Hubble Space Telescope was necessary to determine the Hubble constant, and thus the scale of the Universe, so too did the astronomers of the 18$^{th}$ century argue for an expedition to measure the transit of Venus and thus to determine the scale of the Solar System. 

Since the 1761 transit was poorly observed due to bad weather, in 1767 the Royal Society established a committee to ``report on places where it would be advisable to make the observations, the methods to be pursued, and the persons best suited to carry out the work'' of observing the 1769 transit. After a period of deliberation, the Royal Society, acting much like today's National Academy of Science,  forwarded a proposal to King George III calling for an expedition to Tahiti to observe the transit. The proposal offered a  remarkably modern set of arguments (Hough, 1994, p. 44):

\begin{itemize}

\item It highlighted the  practical applications of the results, noting that transit measurement would ``contribute greatly to the improvement of astronomy on which Navigation so much depends.''

\item It noted with concern potential international competition: ``The French, Spaniards, Danes and Swedes are making the proper dispositions for the Observation thereof...The Empress of Russia has given directions for having the same observed...It would cast dishonor [on the British nation] should they neglect to have the correct observations made of this important phenomenon.''

\item And finally, the proposal included an estimate of the cost of the expedition, exclusive of launch vehicle, that would prove to be an underestimate by a thoroughly modern factor of $\pi$. ``The expense would amount to {\it \char'44}4,000, exclusive of the expense of the ship. The Royal Society is in no condition to defray the expense.''

\end{itemize}

King George approved the project. The Royal Navy provided the  launch vehicle, the collier {\it Endeavour}, and Lieutenant James Cook. The Royal Society provided an observer, Charles Green, and the gentleman-naturalist Joseph Banks, who also contributed more than {\it \char'44}10,000 of his funds. On August 26, 1768, Cook set off with 94 passengers and crew and supplies for 18 months including 604 gallons of rum and 4 tons of beer. They arrived in Tahiti one year later, six weeks before the transit, and set up an observatory at ``Point Venus.'' The observations were successfully  made on June 3, 1769.  The rest of the voyage was not without event as Cook undertook to fulfill the second (and secret) part of his Admiralty orders, namely to search for the mysterious southern continent. He circumnavigated New Zealand's North and South Islands and explored the East Coast of Australia from Cape Hicks, past Sydney and Botany Bay up to the Great Barrier Reef. On sailing up the Whitsunday Passage, just a few miles from Hamilton Island where we now sit, Cook noted ``This land is diversified by hill and valley, wood and lawn, with a green pleasant experience.'' A few days later, however, he ran aground on the reef, which he termed an ``Insane Labyrinth'', nearly sinking before limping to shore for a month of repairs.\footnote{The underwater dangers of Australian waters can still catch the unwary mariner, as the near-sinking of Her Majesty's destroyer {\it HMS Nottingham} after hitting Wolf Rock off the coast of New South Wales during the week of this conference vividly demonstrated.}

Cook returned to England in 1771 with transit data which, when combined with measurements from other sites, led to the determination of the Astronomical Unit to within 3\% of the modern value, a major advance in observational cosmology (Sellers 2001). 

\section{Scientific Introduction}

Two hundred years after Cook's voyage, scientists have started to consider the challenge of finding life on planets beyond our own. As summarized in Woolf and Angel (1998) and Beichman et al. (1999, 2000, and 2002), modern technology offers a realistic opportunity to address this ancient question. In March, 2000, the Terrestrial Planet Finder (TPF) project at JPL selected four university-industry teams to examine a broad range of instrument architectures capable of directly detecting radiation from terrestrial planets orbiting nearby stars, characterizing their surfaces and atmospheres, and searching for signs of life. Over the course of two years the four teams, incorporating more than 115 scientists from 50 institutions worked with more than 20 aerospace and engineering firms. In the first year of study, the contractors and the TPF Science Working Group (TPF-SWG) examined over 60 different ideas for planet detection. Four main concepts, including a number of variants, were selected for more detailed study. Of these concepts, two broad architectural classes appear sufficiently realistic to the TPF-SWG, to an independent Technology Review Board, and to the TPF project that further technological development is warranted in support of a new start around 2010. {\it The primary conclusion from the effort of the past two years is that with suitable technology investment, starting now, a mission to detect terrestrial planets around nearby stars could be launched within a decade.}

The detection of Earth-like planets will not be easy. The targets are faint and located close to parent stars that are $>$1 million (in the infrared) to $>$1 billion times (in the visible) brighter than the planets. However, the detection problem is well defined and can be solved using technologies that can be developed within the next decade. We have identified two paths to the TPF goal of finding and characterizing planets around 150 stars out to distances of about 15 pc: 

$\bullet$ 	At visible wavelengths, a large telescope (a 4x10 m elliptical aperture in one design and an 8x8 m square aperture in another) equipped with a selection of advanced optics to reject scattered and diffracted starlight (apodizing pupil masks, coronagraphic stops, and deformable mirrors) offers the prospect of directly detecting reflected light from Earths. 

$\bullet$	At mid-IR wavelengths, nulling interferometer designs utilizing from three to five 3$\sim$4 m telescopes located on either separated spacecraft or a large, 40 m boom can directly detect the thermal radiation emitted by Earths. 

The TPF-SWG established that observations in either the optical/near-infrared or thermal infrared wavelength region would provide important information on the physical characteristics of any detected planets, including credible signposts of life. In fact, the two wavelengths provide complementary information so that in the long run, both would be desirable. The choice of wavelength regime for TPF will, in the estimation of the TPF-SWG, be driven by the technological readiness of a particular technique. 

\section{Scientific Goals for The Terrestrial Planet Finder}

The TPF Science Working Group (TPF-SWG) established a Design Reference Program to give broad guidelines for defining architectures for TPF. The goals for TPF were set out at the December, 2000, meeting of the TPF-SWG:

{\it Primary Goal for the Terrestrial Planet Finder (TPF)}: TPF must detect radiation from any Earth-like planets located in the habitable zones surrounding ~150 solar type (spectral types F, G, and K) stars. TPF must: 1) characterize the orbital and physical properties of all detected planets to assess their habitability; and 2) characterize the atmospheres and search for potential biomarkers among the brightest Earth-like candidates.

{\it The Broader Scientific Context}: Our understanding of the properties of terrestrial planets will be scientifically most valuable within a broader framework that includes the properties of all planetary system constituents, e.g. both gas giant and terrestrial planets, and debris disks. Some of this information, such as the properties of debris disks and the masses and orbital properties of gas giant planets, will become available with currently planned space or ground-based facilities. However, the spectral characterization of most giant planets will require observations with TPF. TPF's ability to carry out a program of comparative planetology across a range of planetary masses and orbital locations in a large number of new solar systems is by itself an important scientific motivation for the mission. 

{\it Astrophysics with TPF}: An observatory with the power to detect an Earth orbiting a nearby star will be able to collect important new data on many targets of general astrophysical interest. Architectural studies should address both the range of problems and the fundamental new insights that would be enabled with a particular design.

\section{Biomarkers for TPF}

Early TPF-SWG discussions made it apparent that observations in either the visible or mid-infrared portions of the spectrum were technically feasible and scientifically important. A sub-committee of the TPF-SWG was established under the leadership of Dave Des Marais to address the wavelength regimes for TPF. The conclusions of their report (Des Marais {\it et al.} 2002) can be summarized briefly as follows:

$\bullet$ Photometry and spectroscopy in either the visible or mid-IR region would give compelling information on the physical properties of planets as well as on the presence and composition of an atmosphere. 

$\bullet$ The presence of molecular oxygen (O$_2$) or its photolytic by-product ozone (O$_3$) are the most robust indicators of photosynthetic life on a planet. Even though H$_2$O is not a bio-indicator, its presence in liquid form on a planet's surface is considered essential to life and is thus a good signpost of habitability. 

$\bullet$ Species such as H$_2$O, CO, CH$_4$, and O$_2$ may be present in visible-light spectra (0.7 to 1.0 $\mu$m minimum and 0.5 to 1.1 $\mu$m preferred) of Earth-like planets. An ozone band at 0.3 $\mu$m and a general rise in albedo due to Rayleigh scattering are among the few features in the blue-UV part of the spectrum. The lines of these species can be resolved with spectral resolving powers of $\lambda/\Delta\lambda\sim 25-100$.

$\bullet$ Species such as H$_2$O, CO$_2$, CH$_4$, and O$_3$ may be present in mid-infrared spectra of Earth-like planets (8.5 to 20 $\mu$m minimum and 7 to 25 $\mu$m preferred). These lines (except CH$_4$) can be resolved with spectral resolving powers of $\lambda/\Delta\lambda\sim 5-25$.

$\bullet$ The influence of clouds, surface properties (including the presence of photosynthetic pigments such as chlorophyll), rotation, etc. can have profound effects on the photometric and spectroscopic appearance of planets and must be carefully addressed with theoretical studies in the coming years  (e.g. Ford, Seager, and Turner 2001).

In conclusion, the TPF-SWG agreed that either wavelength region would provide important information on the nature of detected planets and that the choice between wavelengths should be driven by technical considerations. 

\section {TPF Architectural Studies}

After an initial year during which the four study teams investigated more than 60 designs, the teams plus JPL identified four architectural classes (with a number of variants) worthy of more intensive study. High level descriptions of these architectures are given below; more detailed information is available in the summary of the recent architecture studies (Beichman {\it et al.} 2002), the final reports from the teams, and, for the separated spacecraft interferometer, the TPF Book (1999). 

\subsection{Visible Light Coronagraphs}

Two groups (Ball and Boeing-SVS) investigated the potential for a visible light coronagraph to satisfy TPF's goals. While there are differences between the designs, there are major similarities: 1) a large optical surface (4 $\times$ 10 m for Ball, 8$\times$8 m for Boeing-SVS); 2) a highly precise, lightweight primary mirror equipped with actuators for figure control with surface quality of order 1-5 nm depending on spatial frequency; and 3) a variety of pupil masks (square, Gaussian, or other (Spergel and Kasdin 2001) and/or Lyot stops) to suppress diffracted starlight. In the case of the Ball designs, a key component was a small deformable mirror with $\sim 100\times 100 = 10^4$ elements capable of correcting residual mid-spatial frequency errors to $\lambda/3,000$ and stable to $\lambda/10,000$. In the Ball design, the combination of pupil masks and the deformable mirror reduces the ratio of starlight (scattered or diffracted)to planet light to approximately unity over an angular extent between $\sim 5 \lambda/D$  and $100 \lambda/D$.

With these features, the Ball systems are able to conduct a survey of 150 stars with images taken at 3 epochs for confirmation and orbital determination in less than half a year. The Boeing-SVS system, as proposed, takes more time to complete such a survey because without a deformable mirror the ratio of starlight to planet light is about 100 times worse than in the Ball design; addition of a deformable mirror would result in comparable performance for the two telescopes. In under a day per star, the Ball system could detect (SNR=5 at spectral resolution $R=\lambda/\Delta\lambda\sim 25-75$) various atmospheric tracers, including O$_2$, a critical signpost for the presence of photosynthetic life. 

The study teams pointed out that the potential for ancillary science was particularly impressive for the visible systems, since it would be straight-forward to add a complement of traditional astronomical instruments, e.g. UV-optical imagers and spectrographs. Operated on an 8-10 m telescope, such instruments would represent a giant advance over the present UV-optical performance of the Hubble Space Telescope. Of particular interest would be the ability to make diffraction limited images at UV-wavelengths with $<5$ milli-arcsec resolution. 

Future studies will have to assess whether the specialized requirements of a planet finding, e.g. an off-axis secondary, might compromise the general astrophysics potential of a visible/ultra-violet system. Conversely, NASA will have to weigh whether specialized needs such as high UV throughput requiring special coatings and careful attention to contamination issues might significantly increase the cost of the observatory or compromise its planet-finding performance.

The greatest technical risk for the visible coronagraph is in the development, manufacturing and implementation of a large primary mirror with ultra-low wavefront errors as well as components associated with starlight suppression. The coronagraphs themselves are functionally simple and although the demands for system performance are challenging, none are thought to be insurmountable. However, the problem of fabricating and launching a large (8-10 m) mirror cannot be overemphasized. The TPF Project's independent Technology Review Board noted that there exists no capability to fabricate such a high precision (3-5 times better than Hubble's mirror, 5-10 times better than NGST's mirror), lightweight optical element for ground or space. But even if a 8-10 m system proves to be too difficult to implement on the TPF timetable, a 2-4 m class telescope could demonstrate high dynamic range coronagraphic imaging and carry out an exciting scientific program. Such a system could find Earths only around the closest dozen stars because of its degraded angular resolution, but it could find and characterize Jupiters around many more distant stars. A telescope of this scale might fit into budget of a Discovery mission.

\subsection{ Nulling Infrared Interferometers}

Lockheed Martin and JPL examined two versions of the infrared nulling interferometer: structurally connected and separated spacecraft. The Lockheed Martin study concluded that a structurally connected infrared interferometer with four 3.5 m diameter telescopes on a fixed 40 m baseline comes close to achieving TPF's goals. The system uses four collinear telescopes arranged as two interleaved Bracewell nulling interferometers to reject star light adequately so that stellar leakage does not compromise the overall system noise. The array would be rotated around the line of sight to the star over a 6-8 hour period. The telescopes can be combined in different pairs to achieve the short and long baselines needed to observe distant or nearby stars. The nulled outputs of the combined pairs are combined again to yield a $\theta^4$ null or an effective $\theta^3$ null with phase chopping.\footnote{In a nulling interferomter, the rejection of starlight is proportional to $\theta^n$, where $\theta \propto (\lambda /B) \phi$,  with $B$ the interferometer baseline and $\phi$ the off-axis angle away from the star. Higher powers of $n$ require more complex optical configurations, but yield broader, deeper nulls.}

The separated spacecraft version of the nulling interferometer was described in the 1999 TPF report (Beichman {\it et al.} 1999; also, see Woolf and Angel 1998). It uses a different arrangement of telescopes to produce a deeper, $\theta^6$, null that can be tuned to resolve most effectively the habitable zone around each target. Because the stellar leakage is reduced in this design, the stability requirements are relaxed relative to the structurally connected interferometer. The $\theta^6$ null is, however, less efficient in its use of baseline, requiring roughly 1.5-2 times longer baselines than the structurally connected system. Providing a 80-100 m baseline leads, in turn, to the likely requirement for a more complex separated spacecraft system. Thus, a near-term study must investigate whether a 40 m system can satisfy TPF's goals. If not, then NASA should pursue aggressively the development of a separated spacecraft nulling interferometer.

It should be mentioned that the European Space Agency (ESA) has studied a two dimensional, separated spacecraft array of infrared telescopes for its Darwin mission. An industrial study by Alcatel found that this version of a planet-finding mission was technically feasible.

The ancillary science possible with an interferometer is likely to be more specialized than for a 8-10 m visible telescope equipped with general purpose instruments. However, the prospect of a telescope with NGST-like sensitivity, but with 10 $\times$ better angular resolution, imaging the cores of protostars, active galaxies, and high redshift quasars is an exciting one.

The largest area of technical risk for the infrared interferometers is not in the performance of the individual components but in the operation of the various elements as a complete system. Most of the required elements are either under development and making good progress, or are reasonable extensions of technology being developed for missions and ground observatories that will be in place well before TPF needs them. However, the system complexity of the separated spacecraft system cannot be overemphasized. Such a system would demand at least one precursor space mission: a formation flying interferometer, such as the Starlight project, to validate the complex control algorithms and beam transport needed for this version of TPF.

\section{Terrestrial Planet Finder (TPF) and the Search for Extraterrestrial Intelligence (SETI) }

This conference offers a happy opportunity for researchers pursuing two quite different techniques for finding extraterrestrial life to come together. NASA's Origins program has focused on a search for planets and primitive  life, in part because of political considerations that ended NASA's involvement in SETI over a decade ago. We are fortunate that dedicated scientists like Jill Tarter and Frank Drake, as well as far-seeing donors such as Paul Allen, have continued to pursue SETI in the context of the vibrant research activities we heard described at this conference.  Despite the political barriers between NASA's programs and SETI, the unity of these efforts can be seen through an examination of the Drake equation

$$ N= (Star\ Formation\ Rate) \times f_{solar\ type\ stars} \times f_{planets}  \times N_\oplus  $$
$$ \ \ \ \ \ \ \ \ \ \  \times  f_{life} \times  f_{intelligence} \times  f_{communicative} \times  Lifetime$$

Twentieth century astronomers determined the first two terms, the star formation rate in the galaxy and the fraction of stars that are of solar type. Present day radial velocity studies combined with future transit experiments (the Kepler mission; Borucki et al. 2001) and astrometric observations (the Space Interferometric Mission, SIM) will determine $f_{planets}$, the fraction of stars with planets,  and $N_\oplus$, the number of earth-like planets in the habitable zone, both statistically and around our nearest neighboring stars.  TPF will characterize the Earth-like planets, habitable or not, and provide the first measurements (or upper limits) of $f_{life}$, the fraction of suitable planets that develop life. Thus, within a generation, we will have well established values for the next three terms in the Drake equation. 

However, astronomical observations cannot determine the final three terms in the equation: the fraction of planets that develop intelligent life, the fraction of intelligent civilizations that can (or want) to communicate, and the lifetime of a communicative civilization. These terms are the realm of the SETI. However, it is the basic frustration of SETI that we cannot separate these remaining terms without at least one successful contact.  This inability to interpret a negative result makes SETI an inherently non-scientific experiment despite its highly technological nature. On the other hand, SETI is an important program of {\it exploration} that must be carried out with our best technology and with great scientific rigor in light of the importance of a positive outcome.  Despite the interesting papers presented at this conference, useful information on the last terms in the Drake equation will not come from terrestrial analogy. Such attempts remind one of Capt. Cook's critics who tried to deduce the existence of a grand Southern Continent by analogy with the Northern Hemisphere (Hough 1994, p. 220,306). It took Cook's explorations to rebut these claims and discover the truth. Similarly, it will take rigorous searches (or serendipitous discovery) to determine whether there is another planet with {\it intelligent}  life in the Universe.

\section{Conclusions}

NASA, together with its potential international partners, has begun to address the challenge of looking for habitable planets and seeking signs of life beyond the Solar System. Captain Cook's voyages remind us that we have asked these questions before and, after much hard work, have been rewarded with new continents to explore.  As Halley said in his paper predicting the 1761 and 1769 transits of Venus (Halley 1716; quoted in Sellers 2001):

\begin{quote}

{\it ``We therefore recommend again and again, to the curious investigators of the stars to whom, when our lives are over, these observations are entrusted, that they, mindful of our advice, apply themselves to the undertaking of these observations vigorously. And for them we desire and pray for all good luck, especially that they be not deprived of this coveted spectacle by the unfortunate obscuration of cloudy heavens, and that the immensities of the celestial spheres, compelled to more precise boundaries, may at last yield to their glory and eternal fame.''}

 \end{quote}

Cook's voyage, in the service of science and exploration, resonates with us today as we use transits of extra-solar planets and other techniques to search for new worlds and for life beyond Earth.  That Cook, Banks, and the English society that dispatched them had multiple motivations for  making this voyage --- some noble, some base --- only highlights the similarities with modern exploration where personal ambition, local politics, and practical applications mix with the search for scientific truth. Today's scientists and policy makers would do well to consider the linkages between science and exploration that future voyages of discovery may enable. The technologies needed to look for other habitable worlds are within our grasp. We need only the will (and the funding) to undertake the search.

\section{Acknowledgements}

The research described in this paper was performed at the Jet Propulsion Laboratory, California Institute of Technology, under a contract with  the National Aeronautics and Space Administration. This work was supported  by the TPF project. The author acknowledges valuable contributions  from TPF Science Working Group and the contractor teams,  as well as the dedication and hard work of Dan Coulter and Chris Lindensmith at JPL. The hospitality and dedication of the conference organizers and beauty of the conference site were appreciated by all the participants.


\begin{references}

\reference{Beichman, C.A., {\it Planetary Systems in the Universe}, International Astronomical Union. Symposium no. 202. Manchester, England (August 2000).}

\reference{Borucki, W. J., Koch, D. G., Jenkins, J. M. 2001, {\it BAAS}, {\bf 199}, 115.04.}

\reference{Des Marais, D., {\it et al.} 2002, {\it Astrobiology}, in press.}

\reference{Ford, E. B., Seager, S., Turner, E. L. 2001, {\it Nature}, {\bf 412},885.}

\reference{Halley, E. 1716, {\it Philosophical Transactions}, XXIX,
{\it A new Method of determining the Parallax of the Sun,
or his Distance from the Earth}, Sec. R. S. NO 348, p.454.}

\reference{Hough, Richard, {\it Captain James Cook: A Biography}, (New York:  W. W. Norton), 1997.}

\reference{Sellers, David, {\it The Transit of Venus}, 2001, (London: Magavelda Press).}

\reference{Spergel, D. and Kasdin, J. 2001, {\it BAAS}, {\bf 199}, 86.03.}

\reference{\textit{Summary Report On Architectural Studies For
The Terrestrial Planet Finder}, 2002, edited by Beichman, C. A., 
Coulter, D., Lindensmith, C and Lawson P. JPL Report, 02-11.}

\reference{\textit{The Terrestrial Planet Finder (TPF): A NASA Origins Program to Search for Terrestrial Planets}, 1999, edited by Beichman, 
C. A., Woolf, N. J., Lindensmith. C. A., JPL Report 99-3.}

\reference{Woolf, N. and Angel, J. R. 1998, {\it ARAA}, {\bf 36}, 507.}


\end{references}
\end{document}